# Medición estandarizada (SMA) versus modelos avanzados (AMA): una revisión crítica de enfoques en Riesgo Operacional

Omar Briceño [0000-0002-6608-4388]

National University of San Marcos, Lima, Perú
omar.briceno@adriskserv.com

**Abstract.** El Comité de Supervisión Bancaria de Basilea propuso reemplazar todos los enfoques para el capital de riesgo operacional, incluido el Enfoque de Medición Avanzada (AMA), por una fórmula simplificada denominada Enfoque de Medición Estandarizada (SMA). Este artículo examina y critica las debilidades y fallos de SMA, como la inestabilidad, la insensibilidad al riesgo, la super-aditividad y la relación implícita entre el modelo de capital de SMA y el riesgo sistémico en el sector bancario. Además, se analizan los problemas del modelo de Capital en Riesgo Operacional (OpCar) propuesto por el Comité de Basilea, precursor de SMA. Se concluye abogando por mantener el marco del modelo interno de AMA y se sugieren una serie de recomendaciones de estandarización para unificar la modelización interna del riesgo operacional. Los hallazgos y puntos de vista presentados en este artículo han sido discutidos y apoyados por numerosos profesionales y académicos de Riesgo Operacional en diversas regiones del mundo.

# 1 Introducción

La gestión del riesgo operacional es la que menos años tiene respecto de las tres ramas principales de riesgo en las instituciones financieras, siendo las otras dos el riesgo de mercado y el riesgo de crédito. El término riesgo operacional se popularizó después de la quiebra del banco Barings en 1995, cuando un operador deshonesto causó el colapso de la institución al realizar apuestas en los mercados asiáticos y mantener estos contratos fuera del alcance de la gestión. En ese momento, estas pérdidas no podían clasificarse ni como riesgos de mercado ni como riesgos de crédito, y se empezó a utilizar el término riesgo operacional en la industria financiera para definir situaciones en las que podrían surgir tales pérdidas. Al principio, riesgo operacional tenía una definición negativa como "cualquier riesgo que no sea de mercado o de crédito" (Basel Committee on Banking Supervision, 1998), lo cual no era útil para evaluar y gestionar el riesgo operacional.

A medida que la gestión de riesgos comenzó a crecer como disciplina, la regulación también se volvió más compleja para adaptarse a las nuevas herramientas y técnicas. Las instituciones financieras siempre han sido reguladas de una forma u otra debido al riesgo que representan para el sistema financiero. La regulación era principalmente a nivel de cada país y muy desigual, permitiendo arbitrajes. A medida que las instituciones financieras se globalizaron, la necesidad de una regulación más simétrica que pudiera nivelar la supervisión y regulación aumentó mecánicamente en todo el mundo.

Como consecuencia de tales regulaciones, ha habido en algunas áreas de la gestión de riesgos, como el riesgo de mercado y el riesgo de crédito, una convergencia o estandarización gradual de las mejores prácticas que han sido ampliamente adoptadas por las instituciones financieras. En el contexto de la modelización y gestión del riesgo operacional, dicha convergencia de mejores prácticas sigue ocurriendo debido a múltiples factores: pasar de un framework que alberga metodologías cualitativas a uno con metodologías cuantitativas; en distintas jurisdicciones en las que operan las entidades financieras, se han registrado diversas experiencias en la captura y mantenimiento de eventos de pérdida según las categorías de riesgo operacional; y la naturaleza misma del riesgo operacional como una categoría de riesgo relativamente inmadura en comparación con riesgo de mercado y de crédito.

La pregunta que surge es cómo se puede inducir una estandarización de la modelización del riesgo operacional y el cálculo de requerimiento de capital bajo el Pilar I de los acuerdos regulatorios bancarios actuales del Comité de Supervisión Bancaria de Basilea (BCBS, por sus siglas en inglés). Según estos acuerdos, el objetivo básico del trabajo del Comité de Basilea ha sido cerrar las brechas en la cobertura de supervisión internacional en busca de dos principios básicos: que ninguna entidad financiera extranjero debe escapar a la supervisión; y que la supervisión debe ser adecuada. Este segundo punto forma el contexto para las nuevas revisiones propuestas para simplificar los enfoques de modelización del riesgo operacional, presentadas en dos documentos consultivos:

- El Enfoque de Medición Estandarizado (SMA) propuesto en el documento consultivo del Comité de Basilea " Standardised Measurement Approach for operational risk", emitido en marzo de 2016 para comentarios hasta el 3 de junio de 2016 (Basel Committee on Banking Supervision, 2016).
- El modelo de Capital en Riesgo Operacional (OpCaR) propuesto en el documento consultivo del Comité de Basilea " Operational risk –Revisions to the simpler approaches" emitido en octubre de 2014 (Basel Committee on Banking Supervision, 2014).

En la página 1 del documento consultivo se menciona que "A pesar del aumento en el número y la severidad de los eventos de riesgo operacional durante y después de la crisis financiera, los requerimiento de capital para riesgo operacional han permanecido estables o incluso han disminuido para los enfoques estandarizados" (Operational risk –Revisions to the simpler approaches, 2014, p. 1). Por lo tanto, es razonable reconsiderar estas medidas de adecuación de capital y evaluar si necesitan una revisión adicional. Este es el proceso que ha llevado a cabo el Comité de Basilea al preparar, revisar y analizar los diferentes componentes propuestos que se discuten en este trabajo de investigación. Antes de revisar el marco propuesto para el cálculo de capital por SMA, es importante recordar las mejores prácticas actuales en las regulaciones de Basilea.

Bajo el marco regulatorio de Basilea II (Basel Committee on Banking Supervision, 2006), se han sugerido muchos modelos para estimar el requerimiento mínimo de capital por el riesgo operacional. Fundamentalmente, se consideran dos enfoques diferentes: el enfoque de arriba hacia abajo (top-down) y el enfoque de abajo hacia arriba (bottom-up). Un enfoque de arriba hacia abajo cuantifica el riesgo operacional sin intentar identificar explícitamente los eventos o causas de las pérdidas. Puede incluir los modelos de indicadores de riesgo que se basan en una serie de indicadores de exposición al riesgo operacional para rastrear estos riesgos y los modelos de análisis de escenarios y pruebas de estrés que se estiman en función de escenarios hipotéticos. Mientras que un enfoque de abajo hacia arriba cuantifica el riesgo operacional a nivel micro, basándose en eventos internos identificados. Puede incluir modelos de tipo actuarial (conocidos como el Enfoque de Distribución de Pérdidas - LDA) que modelizan la frecuencia y la severidad de las pérdidas por riesgo operacional.

Bajo el marco regulatorio actual para el riesgo operacional, los bancos pueden utilizar varios métodos para calcular el capital por riesgo operacional: el Enfoque de Indicador Básico (BIA), el Enfoque Estandarizado (TSA) y el Enfoque de Medición Avanzada (AMA). En resumen, bajo el BIA y el TSA, el capital se calcula como funciones simples del ingreso bruto (GI), el detalle de estos dos enfoques puede encontrarse en el documento de Basilea II (Basel Committee on Banking Supervision, 2006, p. 144).

Para el uso de AMA, las entidades financieras pueden usar modelos propios o internos para estimar el capital requerido. Estos modelos deben demostrar su precisión en celdas de riesgo específicas según Basilea II, las cuales se dividen en 8 líneas de negocio y 7 tipos de eventos. Este enfoque detallado permite un análisis más granular de los riesgos dentro de la institución financiera. El método más utilizado en AMA es el enfoque de distribución de pérdidas (LDA), que modeliza la frecuencia anual N y la severidad $X_1, X_2, X_3\ldots$ de los eventos de pérdidas por riesgo operacional en cada celda de riesgo:

$$Z = \sum_{j=1}^{d} \sum_{i=1}^{N_j} X_i^{(j)} \qquad (1)$$

El cálculo del capital regulatorio es el valor en riesgo (VaR) al 0.999, que es el cuantil de la distribución para la pérdida anual Z del próximo año:

$$K_{LDA} = VaR_q[Z] \coloneqq \inf\{z \in \mathbb{R} : \Pr[Z > z] \leq 1 - q\}, \qquad q = 0.999 \qquad (2)$$

Siendo la frecuencia y severidad variables que se asumen independientes en cada una de las celdas. La expresión anterior puede reducirse siempre y cuando la pérdida esperada sea utilizada como cobertura mediante las provisiones internas.

Durante la última década, la gestión del riesgo operacional ha evolucionado bajo una estructura basada en modelos, influenciada por los acuerdos de Basilea II y III. Estos acuerdos han buscado asignar el capital de manera más sensible al riesgo, mejorar los requerimientos de divulgación, cuantificar riesgos de crédito, operacional y de mercado con datos y técnicas formales, y alinear el capital económico y regulatorio. A pesar de los avances, se ha observado que los métodos BIA y TSA no estiman correctamente el requerimiento mínimo de capital por riesgo operacional, y que el capital bajo AMA es difícil de comparar entre entidades financieras debido a las diferentes prácticas adoptadas. Para afinar y estandarizar las prácticas de modelización en riesgo operacional, se considera mejorar BIA y TSA y eliminar los modelos avanzados (modelos internos) en favor de un modelo SMA simplificado y que "todos puedan usar".

En este documento de investigación se define formalmente el modelo SMA propuesto por Basilea y se recopilan los resultados de estudios realizados sobre el modelo SMA propuesto.

## 2    Comité de Basilea y la propuesta de SMA

Este apartado desarrolla en forma resumida las simplificaciones propuestas por el Comité de Basilea para los modelos de riesgo operacional, comenzando con una visión general del proceso iniciado con el modelo Capital en Riesgo Operacional (OpCaR) propuesto en el documento consultivo "Operational risk –Revisions to the simpler approaches" (Basel Committee on Banking Supervision, 2014) y finalizando con la versión actual conocida como SMA, propuesta en "Standardised Measurement Approach for operational risk" (Basel Committee on Banking Supervision, 2016).

Se inicia con un comentario aclaratorio sobre la primera ronda de propuestas, importante para los profesionales del sector. En la página 1 de BCBS (2014) se afirma que, a pesar del aumento en la cantidad y severidad de eventos de riesgo operacional durante y después de la crisis financiera, los requisitos de capital para el riesgo operacional se han mantenido estables o incluso han disminuido para los enfoques estandarizados. Esto indica que los enfoques simples existentes para el riesgo operacional, el Enfoque del Indicador Básico (BIA) y el Enfoque Estandarizado (TSA), incluidas sus variantes, no estiman correctamente los requisitos de capital para riesgo operacional de un muchas entidades financieras.

En el mismo documento de Basilea, se reconoce que, existen muchos casos en los que las entidades financieras estarán sub-capitalizadas para grandes eventos de crisis, por ejemplo, la ocurrida en 2008. Por lo tanto, es prudente reconsiderar estos modelos simplificados y buscar mejoras y reformulaciones a la luz de la nueva información disponible tras la crisis de 2008.

Se observa que el BIA y el TSA hacen suposiciones muy simplistas sobre el capital, como que el ingreso bruto puede usarse como un indicador adecuado de la exposición al riesgo operacional y que la exposición al riesgo operacional de una entidad financiera aumenta linealmente en proporción a los ingresos. El documento consultivo (Basel Committee on Banking Supervision, 2014, p. 1) también aborda la premisa de que los enfoques actuales no consideran que la relación entre el tamaño de la entidad financiera y el riesgo operacional no es constante o que la exposición al riesgo operacional aumenta de manera no lineal con el tamaño de la entidad.

Además, debe tenerse en cuenta que los enfoques BIA y TSA no han sido recalibrados desde 2004, lo cual se considera un error significativo, todos los modelos y las calibraciones deberían ser validados o testeados regularmente, y cada normativa debe venir con un plan de revisión. La experiencia ha demostrado que las suposiciones del modelo suelen ser inválidas en un entorno de gestión de riesgos dinámicamente cambiante y no estacionario.

Los dos principales objetivos del modelo OpCaR propuesto en Operational risk – Revisions to the simpler approaches (2014) eran: (i) afinar el indicador proxy del riesgo operacional reemplazando el ingreso bruto (GI) con un indicador superior; y (ii) mejorar la calibración de los coeficientes regulatorios basándose en los resultados del análisis cuantitativo.

Para lograr esto, el Comité de Basilea argumentó, que el modelo desarrollado debería ser lo suficientemente simple para aplicarse con "comparabilidad de resultados en el marco" y "lo suficientemente simple de entender, no excesivamente oneroso de implementar, no debería tener demasiados parámetros para el cálculo por parte de las entidades y no debería depender de los modelos internos". Sin embargo, también afirmaron que dicho nuevo enfoque debería "mostrar una mayor sensibilidad al riesgo" en comparación con los enfoques basados en el ingreso bruto.

Además, el enfoque debe ser único, es decir, "debería calibrarse de acuerdo con el perfil de riesgo operacional de un gran número de entidades de diferentes tamaños y modelos de negocio". Pretender que el enfoque cumpla con la premisa de ser único, no tiene sentido, dado que hay muchas entidades financieras en diferentes jurisdicciones, con diferentes tamaños y distintas prácticas bancarias que no permitiría asegurar una distribución poblacional común para todas las entidades financieras.

En el desarrollo de esta investigación, se proporcionará algunas críticas científicas a varios aspectos técnicos de la estimación y las aproximaciones utilizadas, es importante considerar aún estos aspectos, ya que este modelo es el precursor de la fórmula SMA. Es decir, se asume una única LDA para una entidad financiera y se utiliza una aproximación de pérdida única (Single Loss Approximation - SLA,) para estimar el cuantil 0.999 de la pérdida anual. En el documento consultivo, Basilea explica cómo se ajustaron cuatro distribuciones de severidad diferentes a los datos de muchos bancos y se asumió una distribución de Poisson para la frecuencia. Luego, se utiliza una regresión no lineal para ajustar el requerimiento de capital obtenido (a través de muchas entidades) a distintas combinaciones de variables explicativas de los libros bancarios hasta llegar a la fórmula OpCaR.

El SMA combina el indicador de negocio (BI), que es una aproximación simplificada de la exposición al riesgo operacional a partir de los estados financieros, con ciertos datos específicos sobre las pérdidas operacionales de un banco.

$$K_{SMA}(BI, LC) = \begin{cases} BIC, & si\ Bucket\ 1 \\ 110 + (BIC - 100) * \ln\left(\exp(1) - 1 + \frac{LC}{BIC}\right), & si\ Buckets\ 2-5 \end{cases} \quad (3)$$

Es decir, el cálculo del componente de pérdidas se puede expresar mediante:

$$LC = 7\frac{1}{T}\sum_{t=1}^{T}\sum_{i=1}^{N(t)} X_i + 7\frac{1}{T}\sum_{t=1}^{T}\sum_{i=1}^{N(t)} X_i(t)1_{\{X_i>10\}} + 5\frac{1}{T}\sum_{t=1}^{T}\sum_{i=1}^{N(t)} X_i(t)1_{\{X_i>100\}} \quad (4)$$

donde T = 10 años (o al menos 5 años para bancos que no tienen 10 años de datos de pérdidas de alta calidad en el periodo de transición). Los buckets y el componente de indicador de negocio (BIC) son calculados como:

$$BIC = \begin{cases} 0.11 \times BI, & si\ BI \leq 1000, Bucket\ 1, \\ 110 + 0.15 \times (BI - 1000), & si\ 1000 < BI \leq 3000, Bucket\ 2, \\ 410 + 0.19 \times (BI - 3000), & si\ 3000 < BI \leq 10000, Bucket\ 3, \\ 1740 + 0.23 \times (BI - 10000), & si\ 10000 < BI \leq 30000, Bucket\ 4, \\ 6340 + 0.29 \times (BI - 30000), & si\ BI > 30000, Bucket\ 5. \end{cases} \quad (5)$$

El componente BI está definido como la suma del componente de interés, leasing y dividendo, el componente de servicio y el componente financiero, que es diferente para cada uno de los 5 buckets en función del tamaño de su BI; para el detalle de las fórmulas y los componentes, puede revisar Standardised Measurement Approach for operational risk (2016) . Todos los montos están expresados en millones de Euros.

## 3  Inestabilidad del capital por el uso de SMA

Del análisis efectuado, se observa que SMA no logra cumplir con el objetivo de mantener el capital estable. En esta sección se demuestra el resultado mediante diversos ejemplos que han sido preparados utilizando la severidad Lognormal; la frecuencia de Poisson para los eventos de riesgo operacional; así como otras distribuciones que fueron considerados en el modelo OpCaR del documento de Basilea (Basel Committee on Banking Supervision, 2014).

### 3.1  Algunos ejemplos de la inestabilidad del capital

El modelo utilizado para elaborar los ejemplos de inestabilidad representa un proceso anual de pérdidas por riesgo operacional de las entidades que se han descrito líneas arriba, compuesto por dos procesos genéricos de pérdida: uno de alta frecuencia con bajas pérdidas de severidad, y otro de baja frecuencia con pérdidas de alta severidad, modelizados con distribuciones Poisson-Gamma y Poisson-Lognormal, respectivamente. El indicador de negocio (BI) se mantiene constante en 2 mil millones unidades monetarias, es decir, dentro del intervalo del segundo nivel (Bucket 2) de SMA. Se simularon 1,000 años de datos de pérdidas para las tres entidades (pequeña, mediana y grande) con diferentes configuraciones de parámetros para caracterizar las distintas instituciones. Los resultados muestran que una entidad pequeña tiene una pérdida media anual de decena de millones, una mediana de cientos de millones y una grande bordea los 900 millones UM.

El análisis demuestra que, a medida que las instituciones financieras experimentan fluctuaciones significativas en sus ratios de capital SMA, es posible que el capital de una institución se duplique e incluso triplique de un año a otro. Esto ocurre sin cambios en los parámetros, en el modelo o en la estructura de BI, como se observa en la Figura 1, la volatilidad de los ratios del capital SMA, especialmente en entidades grandes. Este comportamiento resalta la naturaleza inherentemente volátil del enfoque SMA, subrayando la importancia de la prudencia al evaluar el capital regulatorio.

**Tabla 1.** Caso 1: proceso Poisson (10) – Lognormal ($\mu = \{10,12,14\}$, $\sigma = 2.5$) y Poisson (990) – Gamma ($\alpha = 1$, $\beta = \{10^4; 10^5; 5 \times 10^5\}$). Caso 2: proceso Poisson (10) – Lognormal ($\mu = \{10,12,14\}$, $\sigma = 2.8$) y Poisson (990) – Gamma ($\alpha = 1$, $\beta = \{10^4; 10^5; 5 \times 10^5\}$).

| Tamaño de la entidad | Caso 1 Pérdida promedio anual (Millones de UM) | Caso 1 Pérdida anual VaR 99.9 (Millones de UM) | Caso 2 Pérdida promedio anual (Millones de UM) | Caso 2 Pérdida anual VaR 99.9 (Millones de UM) |
|---|---|---|---|---|
| Pequeña | 13 | 164 | 15 | 354 |
| Mediana | 130 | 1,663 | 160 | 3,620 |
| Grande | 713 | 9,049 | 897 | 18,832 |

**Figura 1.** Ratio de capital promedio de SMA a largo plazo. Caso 1: σ = 2.5 (gráfico superior); Caso 2: σ = 2.8 (gráfico inferior); los otros parámetros se especifican en

**Tabla 1**.

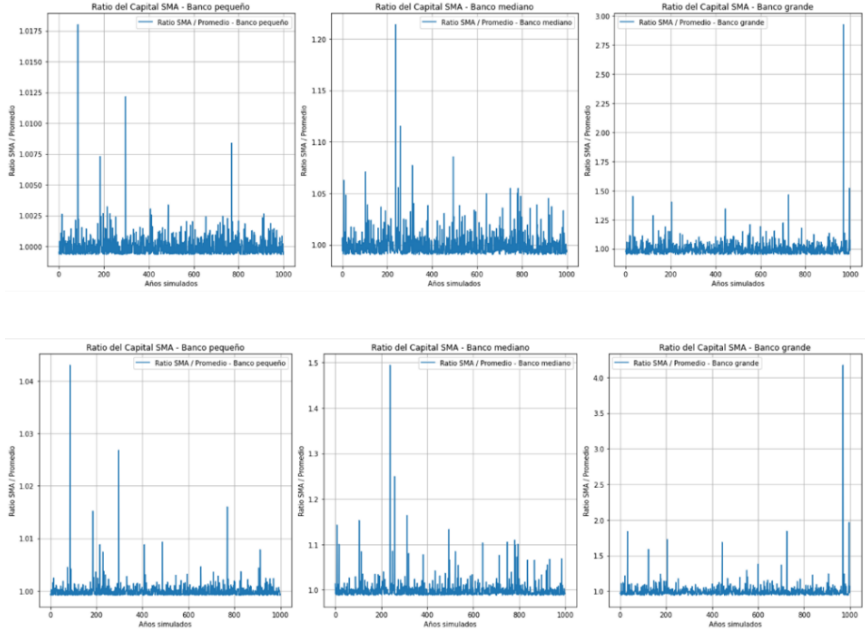

### 3.2 El indicador de negocio (BI) y la inestabilidad del requerimiento de capital al equiparar SMA y AMA

Otro estudio para evaluar la inestabilidad de la carga de capital mediante SMA consistió en modelizar las pérdidas mediante un proceso de Poisson - Lognormal. En lugar de fijar el Indicador de Negocios (BI) en el punto medio del Bucket 2 de la formulación SMA, se ha resuelto numéricamente el valor del BI que produciría un capital SMA igual al Valor en Riesgo (VaR) para un modelo LDA Poisson-Lognormal al cuantil del 99.9% con un horizonte de tiempo anual, es decir, $VaR_{0.999}$.

En otras palabras, se estimó el valor del BI tal que el requerimiento de capital bajo el modelo LDA coincida con el requerimiento de capital de SMA a largo plazo. Esto se logra resolviendo numéricamente la siguiente ecuación no lineal a través de la búsqueda de raíces para el BI:

$$K_{SMA}(BI, \widetilde{LC}) = VaR_{0.999}, \tag{6}$$

donde $\widetilde{LC}$ es el promedio a largo plazo del componente de pérdida (4) que puede ser estimado para el caso de la frecuencia, mediante una distribución de Poisson ($\lambda$):

$$\widetilde{LC} = \lambda \times (7E[X] + 7E[X \backslash X > L] + 5E[X \backslash X > H]). \tag{7}$$

Para el caso de la severidad $X$ desde una distribución Lognormal ($\mu, \sigma$), se estima analíticamente como:

$$\widetilde{LC}(\lambda, \mu, \sigma) = \lambda e^{\mu + \frac{1}{2}\sigma^2} \left( 7 + 7\Phi\left(\frac{\sigma^2 + \mu - \ln L}{\sigma}\right) + 5\Phi\left(\frac{\sigma^2 + \mu - \ln H}{\sigma}\right) \right). \tag{8}$$

Donde $\Phi(.)$ denota la función de la distribución Normal estándar, L es 10 millones de unidades monetarias (UM) y H es 100 millones de UM como se especifica en la fórmula de SMA (4).

El VaR bajo el modelo Poisson - Lognormal según la aproximación de pérdida única o SLA (single loss approximation) para un nivel de confianza $\alpha$ puede escribirse como:

$$\text{VaR}_\alpha \approx \text{SLA}(\alpha; \lambda, \mu, \sigma) = \exp\left(\mu + \sigma\Phi^{-1}\left(1 - \frac{1-\alpha}{\lambda}\right)\right) + \lambda \exp\left(\mu + \frac{1}{2}\sigma^2\right); \tag{9}$$

donde $\Phi^{-1}(.)$ es la inversa de la función de la distribución Normal estándar. En este caso, los resultados para los valores implícitos del BI se presentan en la Tabla 2, con un parámetro $\lambda = 10$ y variando los parámetros de la distribución Lognormal $\mu$ y $\sigma$. Cabe destacar que tampoco es complicado calcular el $VaR_\alpha$ de manera "exacta" (dentro de un margen de error numérico) utilizando métodos numéricos como Monte Carlo, la recursiva de Panjer o la Transformada Rápida de Fourier (FFT).

**Tabla 2.** BI en billones, $\lambda = 10$

| $\mu / \sigma$ | 1.5 | 1.75 | 2 | 2.25 | 2.5 | 2.75 | 3 |
|---|---|---|---|---|---|---|---|
| 10 | **No converge** | | | 0.56 | 2.09 | 3.59 | 9.74 |
| 12 | 0.14 | 1.43 | 4.59 | 7.11 | 15.18 | 29.45 | 136.72 |
| 14 | 4.18 | 7.86 | 38.93 | 27.93 | 88.54 | 338.90 | 350.10 |

La tabla proporcionada refleja el análisis del Indicador de Negocio (BI) para un valor de $\lambda = 10$, variando la relación $\mu/\sigma$ y los valores de $\mu$. Los resultados muestran una tendencia general en la que el BI aumenta significativamente a medida que crece la relación $\mu/\sigma$. Esto implica que, cuando la media de la distribución lognormal ($\mu$) es mucho mayor en relación con la desviación estándar ($\sigma$), el BI presenta incrementos pronunciados. Además, se observa que, en ciertos casos, como cuando $\sigma = 1.5$ y $\mu = 10$, el modelo no converge, lo que puede deberse a una alta dispersión en los datos que genera valores extremos o inestables.

Otro aspecto relevante es el impacto directo de $\mu$ sobre el BI. A medida que este parámetro aumenta de 10 a 14, el BI escala de forma exponencial, especialmente para relaciones $\mu / \sigma$ altas. Por ejemplo, para $\sigma = 2.75$, el BI pasa de 3.59 cuando $\mu = 10$ a 29.45 para $\mu = 12$, y alcanza 338.90 con $\mu = 14$. Este comportamiento evidencia cómo el valor absoluto de µ tiene un efecto multiplicador sobre el BI, generando una fuerte dependencia en los resultados.

El análisis también demuestra un comportamiento no lineal en el crecimiento del BI con respecto a los parámetros de la distribución lognormal. Este efecto se acentúa en valores altos de µ/σ, donde el incremento en el BI es desproporcionado. Por ejemplo, cuando σ = 3 y µ = 14, el BI alcanza 350.10, mostrando un crecimiento exponencial respecto a valores más bajos de µ/σ. Este fenómeno pone de manifiesto la sensibilidad extrema del BI frente a los parámetros de entrada, lo que puede traducirse en variaciones significativas en los cálculos regulatorios.

Estos resultados tienen implicancias importantes para el modelo SMA, ya que demuestran que el BI puede experimentar fluctuaciones extremas sin necesidad de cambiar la estructura del modelo o los parámetros básicos. Esto sugiere que el modelo SMA puede generar resultados altamente volátiles y, en algunos casos, sobreestimar los requerimientos de capital regulatorio. La sensibilidad del BI a pequeñas variaciones en los parámetros de entrada refuerza la crítica de que el modelo no refleja con precisión las condiciones reales de una institución financiera.

### 3.3 SMA es muy sensible a las colas pesadas de una distribución de pérdidas

Para este análisis se ha considerado emular una entidad con un amplio rango de eventos de pérdida en sus diferentes líneas de negocio. Tal cual se hizo en el primer estudio, el proceso de pérdidas se ha dividido en alta frecuencia/baja severidad, y baja frecuencia/alta severidad, modelizado mediante Poisson (990) – Gamma ($1.5 \times 10^5$) y Poisson (10) – Lognormal (14, σ), respectivamente.

La Figura 2 se construyó sobre la base de simulaciones realizadas para distintos valores del parámetro sigma, variando de 2 a 3 (σ = {2, 2.25, 2.50, 2.75, 3}. Cada uno de estos escenarios se simuló a lo largo de 1,000 años, permitiendo observar la distribución del ratio de capital calculado por el SMA bajo diferentes grados de dispersión.

El análisis refuerza que, a medida que sigma aumenta, la dispersión de los resultados crece de manera significativa, lo que se evidencia en el incremento de la amplitud de los boxplots y en el número creciente de valores atípicos (outliers) que alcanzan ratios extremos. Este comportamiento refleja una mayor volatilidad del modelo SMA en entornos con colas más pesadas, característica propia de distribuciones Lognormal cuando sigma es elevado. En particular, el aumento de sigma amplifica las contribuciones de los eventos extremos, lo que se traduce en fluctuaciones más pronunciadas y ratios de capital que, en algunos casos, llegan a quintuplicar el valor central de la mayoría de los años simulados.

La volatilidad en estos escenarios no solo afecta la estabilidad de las estimaciones, sino que también incrementa la incertidumbre en la planeación y provisión del capital regulatorio. Por ejemplo, mientras que para valores de sigma más bajos (como 2.0) los resultados son más concentrados, a partir de sigma 2.75 y 3.0 la dispersión se amplifica dramáticamente, generando valores extremos que distorsionan las métricas generales y representan un reto importante para las instituciones.

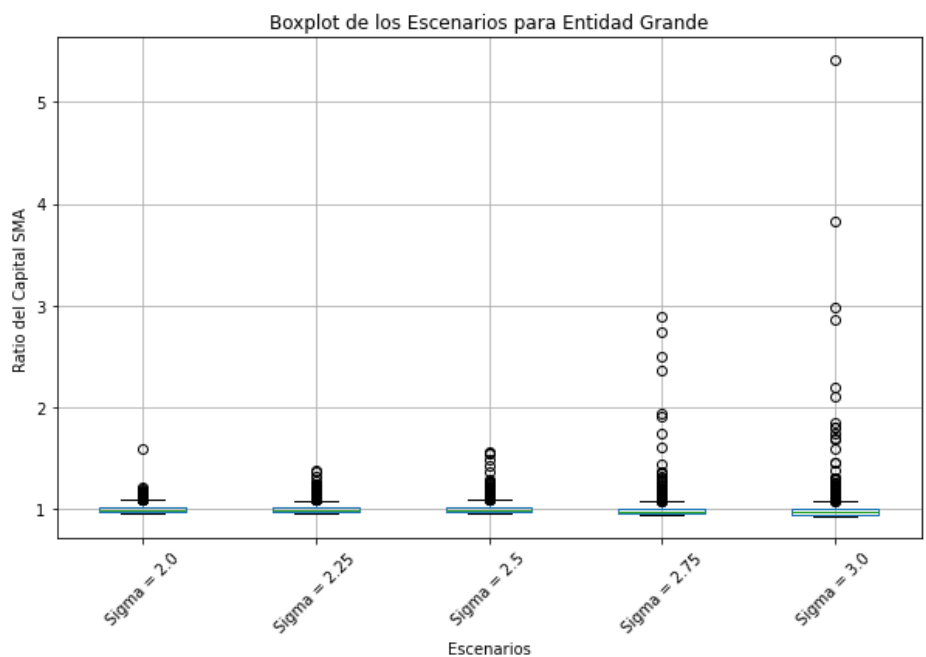

**Figura 2.** Resultados de boxplot para el ratio de capital por SMA a largo plazo variando el parámetro sigma de una distribución Lognormal sobre la base de 1,000 años.

## 4  SMA reduce la sensibilidad al riesgo e induce a la toma de riesgos

El cálculo de capital bajo el enfoque SMA evidencia una marcada falta de capacidad para responder adecuadamente al riesgo en comparación con la granularidad detallada del marco AMA definido por Basilea II, que separaba los eventos de riesgo operacional en 56 celdas. Estas celdas agrupaban líneas de negocio y tipos de eventos de pérdida, proporcionando una visión más detallada de los diferentes perfiles de riesgo dentro de una institución financiera. Bajo AMA, la granularidad permitía a las instituciones adaptar las estimaciones de capital a las características específicas de su exposición al riesgo, logrando una asignación más precisa y alineada con su realidad operativa. En contraste, el enfoque SMA, al depender de un modelo simplificado basado en el Indicador de Negocio (BI), pierde esta capacidad para diferenciar los riesgos y responde de manera limitada a las variaciones en el perfil de riesgo subyacente.

El SMA no está diseñado para mantener una sensibilidad al riesgo adecuada en el cálculo de capital, ya que se basa en parámetros fijos que no incorporan el impacto de las condiciones específicas de cada institución o de los eventos de cola pesada que puedan surgir. Esto genera un incentivo perverso hacia una mayor toma de riesgos ("more risk-taking"), ya que el cálculo de capital no refleja adecuadamente el aumento de la exposición al riesgo operacional. En otras palabras, instituciones con perfiles de riesgo elevados podrían no percibir penalidades proporcionales en términos de capital regulatorio, alentándolas, de forma indirecta, a asumir mayores riesgos.

Además, el SMA carece de mecanismos que incentiven la correcta identificación y reporte de eventos de pérdida. Este fenómeno, conocido como "denying loss events", es problemático porque puede llevar a las instituciones a minimizar la visibilidad de sus pérdidas operacionales para evitar impactos reputacionales o regulatorios, lo que socava la calidad de los datos necesarios para una gestión de riesgos efectiva.

Otro desafío asociado al SMA es la reducción de la actividad de provisiones, conocido como el "hazard of reduced provisioning activity". A diferencia de los requerimientos establecidos bajo IFRS 9, donde las provisiones están directamente relacionadas con la evaluación de pérdidas esperadas y deben actualizarse regularmente, el SMA no integra definiciones claras sobre provisiones en su marco. Esto genera ambigüedad en la provisión y variabilidad en el capital resultante, dado que las instituciones no cuentan con guías claras para alinear sus provisiones al capital requerido. Esta falta de precisión reduce la capacidad de las instituciones para adaptarse dinámicamente a las condiciones cambiantes del mercado y manejar adecuadamente los riesgos de pérdida futura.

En términos de agregación de pérdidas, el SMA agrupa todas las pérdidas en un marco único sin diferenciar su origen o naturaleza. Este agrupamiento elimina la posibilidad de analizar la heterogeneidad de los riesgos operacionales, ocultando información valiosa que podría ser utilizada para mejorar la toma de decisiones y fortalecer las estrategias de mitigación. En el caso de AMA, la separación de riesgos por la línea de negocio y tipo de evento proporcionaba una perspectiva más rica y accionable.

Finalmente, el SMA sufre de una desconexión fundamental con el futuro, al ignorar las proyecciones o escenarios potenciales que podrían alterar los perfiles de riesgo. Mientras que enfoques más avanzados pueden integrar simulaciones y proyecciones en sus cálculos, el SMA permanece anclado en un enfoque retrospectivo que no refleja adecuadamente los riesgos emergentes. Esta falta de proactividad debilita la capacidad de las instituciones para anticiparse a eventos adversos y preparar sus estructuras de capital de manera adecuada.

## 5 La superaditividad de SMA en el requerimiento de capital

El enfoque SMA presenta una característica intrínseca que lo hace susceptible a la superaditividad en el cálculo del capital. Este fenómeno se refiere a la tendencia de SMA de generar estimaciones de capital que son mayores que la suma de los capitales asociados a cada una de las líneas de negocio o categorías de pérdida por separado, en comparación con un enfoque más granular como AMA. Esto se origina debido a su estructura agregada, que omite cualquier consideración explícita de la correlación o independencia entre los componentes individuales de riesgo.

En el contexto del cálculo del VaR (Value at Risk) bajo AMA, se considera una granularidad que abarca múltiples líneas de negocio y tipos de eventos de pérdida. Cada línea de negocio y tipo de evento es modelizada individualmente, con distribuciones específicas que representan tanto la frecuencia como la severidad de las pérdidas. La agregación final utiliza técnicas avanzadas como la convolución de distribuciones, que incorporan explícitamente correlaciones entre las categorías de riesgo. Esto permite obtener un VaR consolidado que refleja de manera más precisa la interacción entre los diferentes componentes del riesgo, evitando sobreestimaciones o subestimaciones significativas.

En contraste, el SMA se basa en una fórmula estandarizada que utiliza como entrada el Indicador de Negocio (BI) y otros parámetros fijos. Este enfoque ignora por completo las diferencias estructurales entre líneas de negocio y tipos de eventos de pérdida, así como las correlaciones existentes entre ellos. Al hacerlo, introduce un grado de agregación excesivo que puede resultar en cálculos de capital que no representan fielmente la naturaleza del riesgo subyacente. En términos matemáticos, esto se traduce en una falta de sensibilidad al riesgo inherente al SMA, lo cual puede derivar en capitales que son superaditivos, es decir, que sobrepasan la suma de los capitales individuales obtenidos bajo AMA.

Además, la superaditividad en el SMA se ve exacerbada en escenarios donde las colas de las distribuciones de pérdida son pesadas, lo cual es una característica común en el riesgo operacional. En estos casos, el SMA no distingue adecuadamente la contribución de eventos extremos, mientras que el AMA, al basarse en distribuciones específicas y enfoques como el Single Loss Approximation (SLA), permite captar con mayor precisión estas colas pesadas en la estimación del VaR.

Por ejemplo, bajo el AMA, una entidad puede calcular el capital requerido considerando una distribución Lognormal para la severidad de las pérdidas y una distribución de Poisson para la frecuencia de eventos. Esto permite incorporar las características particulares de cada línea de negocio y ajustar los resultados según la correlación observada entre categorías. En cambio, el SMA asignará un capital basado en una fórmula genérica que, aunque más sencilla de aplicar, pierde esta capacidad de discriminación y ajuste.

### 5.1 Algunos ejercicios de SMA

En este ejercicio se observa claramente que el enfoque SMA resulta superaditivo. Imaginemos dos entidades con el mismo indicador de negocio (BI) y componente de pérdida (LC). La primera entidad tiene una sola línea de negocio, mientras que la segunda posee dos líneas de negocio, de modo que cada una contribuye con la mitad del BI y del LC total de la primera (Figura 3).

En teoría, al agregar los riesgos de dos líneas de negocio que individualmente representan la mitad de la exposición de una única línea, se debería obtener un capital total igual o inferior a la suma de los capitales calculados de manera individual, gracias a los beneficios de diversificación; es decir, se esperaría que el riesgo agregado fuera menor que la suma de los riesgos individuales (propiedad de subaditividad). Sin embargo, el modelo SMA, al utilizar una fórmula estandarizada basada en el BI y el LC sin capturar adecuadamente las correlaciones y diversificaciones entre las líneas de negocio, termina asignando un capital total que excede la suma de los componentes individuales. Esto significa que, aunque la segunda entidad, dividida en dos líneas, tenga la mitad del

BI y LC de la primera en cada línea, la agregación de ambas líneas produce un capital requerido superior al que se obtendría para la entidad con una sola línea. Dicho resultado evidencia que el SMA es excesivamente sensible al proceso dominante de pérdida y no refleja los beneficios de diversificación, lo que puede llevar a asignaciones de capital más elevadas de lo necesario, en contraposición a lo que se esperaría bajo una medida de riesgo coherente.

Aunque los datos del ejercicio fueron arbitrarios, se llevó a cabo un análisis similar modelizando el BI, el LC y el VaR del enfoque LDA, manteniendo las mismas características para las entidades y sus líneas de negocio. Se asumió una entidad con perfil de riesgo definido mediante una distribución Poisson ($\lambda$) - Lognormal ($\mu$, $\sigma$). Utilizando las ecuaciones (8) y (9), se determinó el valor del BI que se relaciona con el capital SMA al 0.999 del VaR. En el caso de la entidad 2, se aplicó el mismo tratamiento, pero considerando la mitad de los valores del perfil de riesgo para cada línea, es decir, modelizando cada línea con Poisson ($\lambda/2$) - Lognormal ($\mu$, $\sigma$).

En la entidad 1 se utiliza la totalidad de la tasa de eventos ($\lambda=10$), mientras que en la entidad 2, cada línea opera con $\lambda/2$, y luego se agregan las pérdidas de ambas líneas. Con el enfoque AMA, se modelizan de forma detallada la frecuencia y la severidad para cada línea, permitiendo una mayor granularidad y captación de la diversificación entre líneas de negocio. Sin embargo, el modelo SMA, al basarse en parámetros agregados y en el BI, tiende a producir resultados que pueden ser superaditivos, es decir, el capital requerido para la entidad dividida (Entidad 2) puede no ser inferior a la suma de los riesgos de cada línea, o incluso puede ser mayor, si el modelo no captura adecuadamente los beneficios de diversificación.

|  | Componente | Grupo |  |  | Componente | Grupo |  |
|---|---|---|---|---|---|---|---|
| **Entidad 1** | BI | 32,000 |  | **Entidad 1** | BI | 70,000 |  |
|  | BIC | 6,920 |  |  | BIC | 17,940 |  |
|  | LC | 4,000 |  |  | LC | 4,000 |  |
|  | SMA | 5,771 |  |  | SMA | 11,937 |  |

|  | Componente | A | B |  | Componente | A | B |
|---|---|---|---|---|---|---|---|
| **Entidad 2** | BI | 16,000 | 16,000 | **Entidad 2** | BI | 35,000 | 35,000 |
|  | BIC | 3,120 | 3,120 |  | BIC | 7,790 | 7,790 |
|  | LC | 2,000 | 2,000 |  | LC | 2,000 | 2,000 |
|  | SMA | 2,694 | 2,694 |  | SMA | 5,337 | 5,337 |
|  | **SMA Total** | **5,387** |  |  | **SMA Total** | **10,674** |  |

**Figura 3.** Ejercicios realizados para probar la superaditividad (todos los montos expresados en millones de UM). La figura de la izquierda representa a la entidad 1, perteneciente al bucket 5 y las líneas de la entidad 2, pertenecen al bucket 4. La figura de la derecha ambas pertenecen al bucket 5.

Este ejercicio ilustra cómo, al mantener constantes los parámetros $\mu$ y $\sigma$, la agregación de dos líneas (cada una con la mitad de la frecuencia) se comporta de manera diferente en el cálculo de LC y el VaR (ver Figura 4).

La comparación entre la entidad 1 y la entidad 2 permite evidenciar la falta de sensibilidad de SMA ante la diversificación, reforzando las críticas sobre su superaditividad. De esta forma, se puede argumentar que, a pesar de que el enfoque AMA permite capturar la granularidad y responder a los cambios en el riesgo, el SMA podría sobreestimar el capital requerido, especialmente cuando se agrega información de diferentes líneas de negocio.

| | Componente | Grupo | | Componente | A | B |
|---|---|---|---|---|---|---|
| | λ | 10 | | λ | 5 | 5 |
| | μ, | 14 | | μ, | 14 | 14 |
| | σ | 2 | | σ | 2 | 2 |
| **Entidad 1** | BI | 13,960 | **Entidad 2** | BI | 6,980 | 6,980 |
| | BIC | 2,651 | | BIC | 1,166 | 1,166 |
| | LC | 1,321 | | LC | 661 | 661 |
| | SMA | 2,133 | | SMA | 983 | 983 |
| | LDA | 2,133 | | LDA | 1,473 | 1,473 |
| | | | | **SMA Total** | **1,965** | |

**Figura 4.** Ejercicios realizados para probar la superaditividad (todos los montos expresados en millones de UM). BI para la entidad 1 proporciona el perfil de riesgo mediante LDA en función del 0.999 VaR de Poisson (λ) - Lognormal (μ, σ).

Este análisis sienta las bases para profundizar en cómo la estructura simplificada de SMA conduce a una estimación del capital que no aprovecha los beneficios de la diversificación que se observa con AMA.

## 6 Estructura metodológica para la estimación del capital en riesgo operacional (OpCaR)

La estimación del capital en riesgo operacional (OpCaR) se fundamenta en el enfoque de distribución de pérdidas (LDA, por sus siglas en inglés). Este método es ampliamente utilizado en la industria financiera y se basa en modelizar la distribución de pérdidas agregadas a partir de dos componentes: la distribución de frecuencia y la distribución de severidad.

$$Z = \sum_{i=1}^{N} X_i, \tag{10}$$

donde, N es el número anual de pérdidas modelizadas como una variable aleatoria desde una distribución de Poisson, y $X_i$ es la severidad de la pérdida modelizada como una variable aleatoria de distribución $F_x(x;\theta)$ parametrizada por el vector $\boldsymbol{\theta}$. Se asume que N y $X_i$ son independientes. $F_x(x;\theta)$ puede ser modelizado con distribuciones de uno o dos parámetros: Pareto, Lognormal, Log-logística, Log-gamma, Weibull, entre otras.

Briceño (2015) destaca la importancia de modelizar las colas de la distribución de severidad utilizando la Teoría de Valor Extremo (EVT), particularmente el método Picos sobre Umbral (POT). Este enfoque ajusta una Distribución Generalizada de Pareto (GPD) a los datos que exceden un umbral predefinido, lo que permite una estimación más precisa de las pérdidas extremas.

## 6.1 Estimación del capital

El capital regulatorio y económico para el riesgo operacional se estima generalmente a partir del valor en riesgo (VaR) con un nivel de confianza del 99.9%. La obtención del VaR requiere la construcción de la distribución agregada de pérdidas, que puede calcularse con diversos métodos:

- **Panjer recursivo**: Método utilizado cuando la distribución de frecuencia pertenece a la familia de (a, b, 0), permitiendo calcular la convolución de manera eficiente sin recurrir a simulaciones.

- **Transformada rápida de Fourier (FFT)**: Utiliza la transformada de Fourier para realizar la convolución de manera computacionalmente eficiente, transformando las distribuciones al dominio de la frecuencia y luego aplicando la transformada inversa.

- **Monte Carlo**: Método basado en simulaciones aleatorias de eventos de pérdida, permitiendo obtener la distribución agregada a partir de un gran número de escenarios.

En algunos casos, la modelización de severidad requiere el uso de mixturas de distribuciones o la aplicación de teoría de valores extremos (EVT) para capturar adecuadamente la cola de la distribución. Autores como Cruz (2002), Shevchenko (2011) y Chernobai (2008) han propuesto enfoques avanzados para mejorar la estimación del capital considerando eventos extremos y correlaciones entre líneas de negocio.

## 6.2 Selección del modelo

Modelizar el requerimiento de capital en riesgo operacional (OpCar) es un proceso complejo que implica la evaluación de múltiples modelos de severidad. Dado que las pérdidas por riesgo operacional suelen caracterizarse por una alta variabilidad y colas pesadas, es fundamental seleccionar un modelo que capture de manera precisa tanto las pérdidas comunes como las extremas. Sin embargo, no existe un modelo único que se ajuste perfectamente a todos los conjuntos de datos, por lo que es esencial probar, validar y comparar diversas opciones antes de tomar una decisión informada.

Un modelo adecuado no solo debe ajustarse bien a los datos históricos, sino también ser capaz de predecir pérdidas futuras con precisión, especialmente en la cola de la distribución, donde se encuentran las pérdidas más significativas. Esto es particularmente relevante en el contexto regulatorio, donde el capital debe ser suficiente para cubrir pérdidas extremas con un alto nivel de confianza (por ejemplo, 99.9%).

La elección del modelo de severidad debe guiarse por varios criterios:

(1) Ajuste Estadístico. El modelo debe ajustarse bien a los datos históricos, tanto en el cuerpo como en la cola de la distribución. Para evaluar esto, se utilizan pruebas de bondad de ajuste como:

- Kolmogorov-Smirnov (KS): Compara la distribución empírica con la teórica.
- Anderson-Darling (AD): Es más sensible a las discrepancias en las colas.
- Criterios de Información (AIC, BIC): Penalizan la complejidad del modelo, favoreciendo aquellos que equilibran ajuste y parsimonia.

(2) Flexibilidad. El modelo debe ser lo suficientemente flexible para adaptarse a diferentes tipos de datos y escenarios. Por ejemplo, una mixtura de distribuciones (como Lognormal + Pareto) puede ser útil para capturar tanto pérdidas comunes como extremas.

(3) Validación de Colas. Dado que las pérdidas extremas son críticas para el cálculo del capital regulatorio, es esencial validar el ajuste en las colas de la distribución. Esto puede hacerse mediante:
- Gráficos Q-Q.
- Teoría de Valor Extremo (EVT).

(4) Simplicidad e interpretabilidad. Aunque los modelos complejos pueden ofrecer un mejor ajuste, también pueden ser difíciles de interpretar e implementar. Por lo tanto, es importante encontrar un equilibrio entre precisión y simplicidad.

El enfoque final debe equilibrar precisión y practicidad, asegurando que el modelo seleccionado no solo cumpla con los requerimientos regulatorios, sino que también proporcione una visión útil para la gestión del riesgo operacional.

La estimación del capital en riesgo operacional (OpCar) no se limita únicamente a la modelización de las pérdidas históricas, ya que un enfoque integral debe incorporar también otras fuentes de información clave, como el ***Business Environment and Internal Control Factors (BEICF)*** y los riesgos identificados en el ***Risk Control Self-Assessment (RCSA)***. El BEICF permite evaluar el impacto del entorno operativo y la eficacia de los controles internos, ajustando la exposición al riesgo mediante factores cualitativos que capturan riesgos emergentes o cambios en el entorno que aún no se reflejan en los datos históricos. Por otro lado, el RCSA proporciona una visión proactiva de los riesgos potenciales, identificando y priorizando aquellos que pueden no estar representados en las pérdidas históricas debido a su baja frecuencia o a la efectividad de los controles. Ambos componentes son esenciales para cumplir con los requisitos regulatorios del Comité de Basilea, que exige considerar tanto factores cuantitativos como cualitativos en la estimación del capital. Sin embargo, en el Enfoque de Medición Estandarizada (SMA) propuesto por Basilea, estos dos aspectos no están incluidos, lo que hace que el cálculo del capital sea incompleto y menos adaptativo a las particularidades de cada institución.

# 7 Mejor estandaricemos los modelos avanzados - AMA

A la luz de lo expuesto en los apartados anteriores, se debe reconsiderar la idea de que el SMA haya sido el reemplazo definitivo de los enfoques avanzados (AMA). AMA no debió ser descartado**, ¡debió ser estandarizado!** mediante un riguroso proceso estadístico que definiera con precisión las técnicas y métricas a emplear. En lugar de suprimir un enfoque basado en el perfil de riesgo de cada institución, se pudo haber establecido un marco normativo que garantizara su comparabilidad y supervisión efectiva, sin comprometer su capacidad predictiva y adaptabilidad.

Para modelar el requerimiento de capital por riesgo operacional utilizando AMA, el regulador podría proponer una estandarización de este enfoque basada en el conocimiento histórico de las pérdidas. Esta estandarización incorporaría los elementos clave, como los ***Business Environment and Internal Control Factors (BEICF)***, los riesgos identificados en el ***Risk Control Self-Assessment (RCSA)***, así como ***datos externos*** y ***escenarios de estrés***. De este modo, se conservaría la riqueza metodológica de AMA, garantizando al mismo tiempo su coherencia regulatoria y su aplicabilidad en distintas instituciones.

En términos de modelización, se podría establecer un marco común que utilice una distribución de Poisson para la frecuencia de eventos y una distribución Gamma Generalizada para la severidad de las pérdidas. La elección de la Gamma Generalizada permitiría capturar de manera flexible los principales modelos de distribución de pérdidas observados en las últimas dos décadas de gestión del riesgo operacional, incluyendo Weibull, Lognormal, Gamma y Pareto.

Como parte de la propuesta de estandarización de AMA, se recomienda el uso de *Generalized Additive Models for Location, Scale and Shape* – **GAMLSS** – (Rigby y Stasinopoulos, 2010) para mejorar el tratamiento de las líneas de negocio (BL) y los tipos de eventos (ET) en riesgo operacional, ya que estos elementos determinan tanto la frecuencia como la severidad en la metodología Loss Distribution Approach (LDA). La flexibilidad de GAMLSS permitiría modelizar de manera más precisa la distribución de pérdidas, capturando mejor la influencia de los BEICF en la estimación del capital regulatorio, lo cual permitiría desarrollar un **LDA híbrido**, combinando estructuras paramétricas y semiparamétricas para una mejor adaptación a las características específicas de cada institución.

Además, se sugiere que el regulador proporcione un conjunto de indicadores para cada línea de negocio (BL) o tipo de evento (ET), los cuales se integren como parte de los factores de regresión de los modelos de LDA. Estos factores no deben ser indicadores individuales, sino cestas de indicadores, agrupadas de acuerdo con su función y relevancia en el análisis del riesgo operacional. Esta metodología permitiría crear familias de indicadores basadas en distintas categorías como indicadores de exposición, indicadores de estrés, indicadores causales e indicadores de fallas. Este enfoque tiene la ventaja de que las entidades no solo comenzarán a incorporar de manera estructurada y estadísticamente rigurosa la información de BEICF mediante GAMLSS, sino que también se verán incentivadas a recopilar estos factores, estableciendo y siguiendo políticas claras y bien definidas.

# 8 Conclusiones

En este documento se ha analizado y discutido en detalle las debilidades y desventajas de la fórmula de la Medida Estandarizada (SMA) para el cálculo de capital por riesgo operacional, tal como fue propuesto por el Comité de Basilea. A través de un estudio exhaustivo, se ha determinado que el SMA presenta problemas significativos en cuanto a su sensibilidad al riesgo, inestabilidad y superaditividad en la estimación del capital, lo cual limita su capacidad para reflejar de manera precisa la exposición al riesgo operacional de las entidades financieras.

A partir de mi experiencia en la modelización del riesgo operacional, propongo una revisión profunda del modelo SMA. En este sentido, sugiero una estandarización del Enfoque de Modelos Avanzados (AMA), lo cual permitiría contar con una metodología más robusta y flexible, capaz de abordar de manera efectiva las deficiencias del SMA.

**Referencias**

1. Basel Committee on Banking Supervision. (1998). Operational Risk Management.
2. Basel Committee on Banking Supervision. (2006). International Convergence of Capital Measurement and Capital Standards. https://www.bis.org/publ/bcbs128.htm
3. Basel Committee on Banking Supervision. (2014). Operational risk –Revisions to the simpler approaches. https://www.bis.org/publ/bcbs291.htm
4. Basel Committee on Banking Supervision. (2016). Standardised Measurement Approach for operational risk. Basel Committee on Banking Supervision. https://www.bis.org/bcbs/publ/d355.htm
5. Briceño, O. (2015). La gestión de riesgo operacional: aplicación práctica para estimar el caplital regulatorio. ADRISK.
6. Chernobai, A. (2008). Operational Risk: A Guide to Basel II Capital Requirements, Models, and Analysis.
7. Cruz, M. (2002). Operational Risk: A Guide to Basel II Capital Requirements, Models, and Analysis.
8. Rigby, B., & Stasinopoulos, M. (2010). A flexible regression approach using GAMLSS in R.
9. Shevchenko, P. (2011). Modelling Operational Risk Using Bayesian Inference.